\documentclass{aa}
\def\msun{M_\odot}
\def\mb{M_{\rm b}}
\def\msol{{\rm M}_\odot}
\def\msolyr{\rm M_\odot/yr}
\def\rms{r_{\rm ms}}
\def\ltot{l_{\rm tot}}
\def\rb{\bar r}
\usepackage{graphicx}
\usepackage{natbib}
\usepackage{amssymb,amsmath}
\usepackage{color}

\begin{document}

\title{Implications of the measured parameters of PSR J1903+0327 for its
progenitor neutron star}
\author{M. Bejger\inst{1} \and M.Fortin\inst{1,2} \and
P. Haensel\inst{1} \and J. L. Zdunik\inst{1}}
\institute{N. Copernicus Astronomical Center, Polish
           Academy of Sciences, Bartycka 18, PL-00-716 Warszawa, Poland
\and
LUTh, UMR 8102 du CNRS, Observatoire de Paris, F-92195 Meudon Cedex, France\\
{\tt bejger@camk.edu.pl, fortin@camk.edu.pl, haensel@camk.edu.pl, jlz@camk.edu.pl}}
\offprints{bejger@camk.edu.pl}
\date{Received 13/06/2011 Accepted 19/10/2011}


\abstract{The millisecond pulsar PSR J1903+0327 rotating at 465 Hz has
the second highest precisely measured mass ($1.67~\msol$) and a
weak surface magnetic field ($\simeq 2\times 10^8$~G). It is located in
the Galactic plane, bound in a highly eccentric ($e=0.44$) orbit in
a binary system with a solar-mass main-sequence star. These observational
findings pose a challenge for the theory of stellar evolution.}
{Using the intrinsic parameters of PSR J1903+0327 evaluated from radio
observations (mass $M$, rotation period $P$, and magnetic field $B$
deduced from $P$ and $\dot{P}$) and a model of spin evolution
during the ''recycling'' phase (spin-up by accretion from a low-mass 
companion lost afterwards) that takes into account the
accretion-induced magnetic field decay, we aim to calculate
the mass of its neutron star progenitor, $M_{\rm i}$, at the onset
of accretion. In addition, we derive constraints on the average accretion 
rate $\dot{M}$ and the pre-accretion magnetic field $B_{\rm i}$. 
We also seek for the imprint of the poorly known equation of state of 
dense matter at supra-nuclear densities on the spin-up tracks and the 
progenitor neutron star.}
{Spin-up is modeled by accretion from a thin magnetized disk, using
the magnetic-torque disk-pulsar coupling model proposed by Klu{\'
z}niak and Rappaport. We adopt an observationally motivated model of
the surface magnetic field dissipation caused by accretion.
We consider three equations of state of dense matter, which are 
consistent with the existence of $2.0~\msol$ neutron star. Orbital
parameters in the accretion disk are obtained using the space-time 
generated by a rotating neutron star within the
framework of general relativity.}
{Constraints on the progenitor neutron star parameters and the
accretion itself are obtained. The minimum average accretion rate
should be higher than $2-8\times 10^{-10}~\msol~{\rm yr}^{-1}$,
the highest lower bound corresponding to the stiffest equation of
state. Allowed $B_{\rm i}$-dependent values of $M_{\rm i}$ 
are within $1.0-1.4~\msol$, much lower than the oversimplified  
but widely used $B\equiv 0$ result, where one gets $M_{\rm i}>1.55~\msol$.}
{The influence of magnetic field in the ''recycling''
process is crucial - it leads to a significant decrease in the 
spin-up rate and higher accreted masses, in comparison to the $B=0$
model. The estimated initial neutron-star mass depends on the assumed 
dense-matter equation of state. We also show that the otherwise 
necessary relativistic corrections to the Newtonian model of Klu{\'zniak} 
and Rappaport, related to the existence of the marginally-stable circular 
orbit, can be neglected in the case of PSR J1903+0327.}


\titlerunning{Neutron star progenitor of PSR J1903+0327}
\authorrunning{Bejger et al.}
\maketitle
%
\section{Introduction}
\label{sect:introd}

The discovery of PSR J1903+0327, which was the first millisecond 
radio pulsar found in a binary system with a $1.03\ \msol$ 
main-sequence (MS) companion in an eccentric ($e=0.44$) 95-day 
orbit, poses a challenge to formation theories of millisecond 
pulsars \citep{Champion2008}. On the one hand, its timing parameters 
- spin frequency $f=465$~Hz and spin period derivative 
$\dot{P}=1.88\times 10^{-20}~{\rm s~ s^{-1}}$ - are nothing unusual 
for millisecond pulsars. On the other hand however, the high 
eccentricity of the orbit and the nature of the companion 
(main-sequence star), as well as its location in the vicinity of the 
Galactic plane make it unique. Pulsar-clock stability and the high 
eccentricity of the orbit enabled the pulsar mass measurement to be 
determined (by means of the Shapiro delay) with formidable 
precision. An analysis of the timing data obtained from 2006 
December through to 2010 January using the 305-m Arecibo radio 
telescope and the 105-m Green Bank Telescope, found that the pulsar 
mass is $1.67\pm 0.021 ~M_\odot$ (at the 99.7\% confidence limit) 
and showed strong evidence that effects other 
than space-time curvature, e.g., stellar winds or tidal forces 
acting on MS companion, are negligible \citep{Freire-Nov2010}.

According to the current theory of neutron-star (NS) evolution, 
millisecond radio pulsars ($P<10~$ms) originate from "radio-dead" 
pulsars via the accretion-caused spin-up in low-mass X-ray binaries 
(LMXB, see \citealt{Alpar1982,Radhakrishnan1982}). This so-called 
''recycling'' process is believed to facilitate the spin-up from the 
initial $\sim 0.1$ Hz frequency to $\sim 500$ Hz in $\sim 10^9$ yrs 
and is associated with accretion of $\sim 0.1~\msun$. This idea was 
corroborated by the detection of millisecond X-ray pulsations in 
LMXBs, which were interpreted as the manifestation of rotating and 
accreting NSs \citep {Wijnands1998}. These millisecond X-ray pulsars 
are thought to become radio millisecond  pulsars after the accretion 
process expires. Radio millisecond pulsars are extremely stable 
rotators, with $\dot{P}\sim 10^{-20} - 10^{-19}~{\rm ss^{-1}}$. The 
surface magnetic fields estimated from the timing properties are 
three - four orders of magnitude weaker than those in normal radio 
pulsars, for which $B\simeq 10^{12}$ G. This is explained either by 
the ''burying'' of the original magnetic field under a layer of 
accreted $\sim 0.1~\msun$ material \citep 
{BisnovatyiKomberg1974,Taam1986,Cumming2001} or/and by the Ohmic 
dissipation of electric currents in the accretion-heated crust \citep
{Romani1990,GeppertUrpin1994}. The ''recycling'' in LMXBs is a 
particularly efficient mechanism in dense stellar systems such as 
globular clusters and is in accordance with specific statistics of 
radio millisecond pulsars \citep{Lorimer2008}. Out of a total of 213 
millisecond pulsars, some 100 (i.e., nearly half) are indeed located 
within binaries, while among the 1850 radio pulsars, only 141 (i.e., 
8\%) are in binaries. Moreover, out of all radio millisecond 
pulsars, more than half (130) are found in 26 Galactic globular 
clusters! Finally, as much as 40\% of the 130 globular-cluster radio 
millisecond pulsars are in binaries.

Efficient spin-up to the kHz frequencies is possible because
the low-mass companion remains for $\sim 10^8-10^9$~yrs in the
post-MS phase, overflowing its Roche lobe and feeding the accretion disk around a NS. Consequently, tidal friction within the extended
companion has ample time to circularize the orbit. Only an additional
dynamic perturbation involving a third star, an occurrence that is 
not so rare within a globular cluster, can make the orbit highly eccentric or even
tear the system apart. This may be the origin of
eccentric binaries as well as some isolated millisecond pulsars,
both in globular clusters and the Galactic disk.

However, PSR J1903+0327 does not fit the above picture because neither
its orbit is circular, nor its companion a post-MS star. It is thus
a first specimen of a new group; constructing a viable scenario of
its formation turns out to be a difficult challenge. We briefly and critically 
review the principal scenarios that have been proposed since the discovery 
of the PSR J1903+0327 binary in Sect. \ref {sect:scenarios}. All but one 
are excluded although the surviving {\it triple system}
scenario is not free of its own problems either \citep
{Freire-Nov2010,Portegies2011}. Nevertheless, it is clear that PSR J1903+0327
was not ''recycled'' by its current MS companion. In the following, 
we focus on the LMXB stage of the pulsar evolution that
certainly preceded the formation of the presently observed binary.
We consider the millisecond-pulsar formation model in order to
(hopefully) deduce the parameters of the pre-accretion progenitor
NS and to place constraints on the poorly known equation of state
(EOS) of dense matter.

We provide a brief introduction to accretion spin-up scenario and 
the dissipation of pulsar magnetic field due to accretion in Sects.~
\ref{sect:spinup} and \ref{sect:decayB}. In Sect. \ref{sect:acc-B}, 
we describe the main elements of the \citet{klurap} magnetic-torque 
model (hereafter referred to as KR) acting on a NS accreting from a 
thin accretion disk. We report on efforts to extend the Newtonian KR 
model to include the effects of the space-time curvature, especially 
to take into account the existence of the marginally-stable circular 
orbit predicted by General Relativity. In our ''recycling'' 
simulations, we use three EOS of dense matter consistent with the 
existence of a $2.0~\msol$ NS \citep{Demorest2010} - these equations 
are briefly described in Sect. \ref {sect:EOS}, while the results of 
our simulations are presented in Sect. \ref{sect:results}. As a 
final result, we place constraints on the average accretion rate 
during the spin-up stage, the initial magnetic field, and the mass 
of the progenitor NS. In Sect. \ref{sect:conclusions}, we summarize 
our main results. An Appendix contains a terse overview of the 
influence of the space-time curvature effects, something that was 
neglected in the original KR model, on the spin-up tracks of PSR 
J1903+0327.

\section{Proposed scenarios for the formation of the PSR J1903+0327 binary}
\label{sect:scenarios}
Three formation scenarios, denoted below as {\it I-III}, were previously 
proposed in the discovery paper \citep{Champion2008}, and the first  
scenario was developed further by \cite{Liu2009}. These three scenarios
were ruled out by new observational data combined with theoretical
modeling and a {\it fourth} scenario was advanced by \citet{Freire-Nov2010} 
and \citet{Portegies2011}. These scenarios, together with their
critique, are briefly discussed below:
\subsection*{I. Rapid rotation at birth}
The pulsar was born spinning
rapidly in a core-collapse supernova, in a binary system in the
Galactic disk, with a MS companion, the supernova kick
making the orbit strongly eccentric \citep{Liu2009}. A spin frequency
of 465 Hz was reached as a consequence of accretion from the fallback disk, which
requires less than $10^3-10^4$~yrs before the accretion (or rather
{\it hyperaccretion} -  the mean rate is $\sim 10^{-4}~\msun~{\rm
yr}^{-1}$) stops and the pulsar becomes an active radio emitter.
This scenario was considered unlikely by \citet{Champion2008}, because of 
the very low surface magnetic field ($B\sim 10^8~$G) implied by the 
measured value of $\dot{P}$. \citet{Liu2009} proposed that the low magnetic 
field value was caused by its dissipation associated with accretion. 
However, they used a model of a magnetic
field dissipation developed originally for slow accretion in an
LMXB, which is not valid for a hyperaccreting newly born NS.
Additional arguments, in particular observational ones, against
this scenario can be found in \citet{Freire-Nov2010} and
\citet{Portegies2011}, which is therefore unacceptable.

\subsection*{II. Hierarchical triple system}
The pulsar was recycled in a LMXB and is currently a member of
a hierarchical triple system containing a white dwarf (inner binary
with the pulsar) and a MS star (outer binary). The high eccentricity of
the pulsar orbit is generated by the Kozai resonance \citep{Kozai1962}
between the inner and outer binaries. The scenario is ruled out by
the precise measurement of the optical spectra of the MS companion
of the pulsar permitting the calculation of the companion radial
velocity, which turned out to be consistent with predictions based
on the orbital parameters determined from the pulsar timing. The
optical data lead to an independent estimate of the mass ratio in
the binary \citep{Freire-Nov2010}. Finally, there was no time
dependence of the eccentricity $e$ in the TEMPO2 timing analysis,
while $e$ growth is expected owing to the Kozai resonance in the
triple hierarchical system \citep{Gopakumar2009}. The scenario
should therefore be ruled out as inconsistent with observations
(see also \citealt{Freire-Nov2010,Portegies2011})

\subsection*{III. Pulsar spun up in a LMXB and then ejected into the
Galactic disk together with a newly captured MS companion}
The pulsar was spun up in an LMXB within a globular cluster and then 
exchanged the evolved companion in a process of interaction with a
MS star in the cluster core, its binary orbit becoming very
eccentric. The kick-off in the three-body interaction was
sufficient to eject the final NS-MS binary out of the globular
cluster and into the Galactic disk. However, measurements of the
proper three-dimensional velocity of the binary allowed to track 
its position back in time \citep{Freire-Nov2010}. The
binary was found to always be within 270 pc from the Galactic disk
and farther than 3 kpc from the Galactic center. Therefore, an
exchange interaction (where an evolved companion in a LMXB is 
exchanged for a MS star), which is probable in a dense stellar 
environment such as a globular cluster or the Galactic center, is so 
unlikely that it leads to the rejection of the scenario 
\citep{Freire-Nov2010}.

\subsection*{IV. Pulsar in a triple (tertiary) system with two MS
stars of different masses, the pulsar being spun-up by accretion
from the evolved, more massive companion that was afterwards
expelled from the tertiary to make it a binary}
This scenario was proposed by \citet{Freire-Nov2010} and \citet{Portegies2011}. 
A starting point is a tertiary composed of a massive star and two MS stars of different masses. The massive star collapses, giving birth to a pulsar, the 
tertiary still being bound. The initially more massive MS star (MS1) evolves
forming a LMXB with the pulsar and spinning it up by accretion.
After a substantial mass loss, MS1 is removed from the system owing 
to the ablation/accretion induced by the interaction with the
millisecond pulsar or ejected from the system by means of the three-body
interactions \citep{Freire-Nov2010}. This leaves a system as the
observed one, with a $1.03\;\msun$ companion still at the MS stage,
in an eccentric orbit. The ablation/accretion removal of the
companion was previously proposed as a mechanism producing {\it
isolated recycled pulsars} in the Galactic disk. Ablation by
the NS radiation indeed operates in the PSR B1957+20 binary
(Black Widow), whose companion mass was reduced to $0.025\msun$
\citep{Fruchter1988-Black-Widow-PSR,KluzniakRST1988}. However,
details of the ablation process are rather uncertain and the
timescale needed to completely vaporize the companion can be longer
than the Hubble time \citep {LevinsonEichler1991-ablation-MSP}.
Alternatively, the expulsion of the lower-mass companion owing to the
chaotic character of a three-body interaction in the triple system is
proposed. We note, however that references quoted in \citet
{Freire-Nov2010} do not seem to provide an accurate explanation 
of PSR J1903+0327: \citet{Hut1984} assumes equal masses for the
three bodies, and \citet{Phillips1993} studies the reflection of the pulsar
radio beam from the swarm of asteroids orbiting the pulsar and
derives an upper bound of $\sim 10^{-4}~\msun$ to the mass of the
asteroid-like material orbiting the Vela pulsar within 1 AU. In our
opinion, the scenario presented by \citet{Freire-Nov2010} is not yet
properly supported by reliable quantitative estimates. In 
\citet{Portegies2011}, both the evolution and the
dynamics of the NS+MS1+MS2 system are studied in detail. Several
competing channels of evolution for the tertiary are compared using
advanced numerical simulations. The available parameter space leading
to the observed PSR J1903+0327+MS2 binary is estimated and the
birth-rate of these binaries in the Galactic disk is found to be
acceptable. \citet{Portegies2011} highlighted the need to (observationally)
find a ,,missing link'' in the scenario: a wide LMXB orbited by a
tertiary low-mass MS star. It is clear that the discovery of such a
system would strongly fortify the triple-star scenario.

Four scenarios, reviewed critically above, have been proposed until now.
Only one of them cannot be immediately ruled out by observations, 
but even this one lacks a solid quantitative basis. However, it is clear 
that PSR J1903+0327 has been ''recycled'' in order to explain its present 
spin period, low surface magnetic field, and mass, which is  
much higher than the average $1.4\ \msun$. From now on, we focus
exclusively on the spin-up phase of the pulsar evolution, to
deduce its pre-accretion parameters.

\section{Spin-up by accretion in LMXBs}
\label{sect:spinup}

We consider a binary consisting of a young radio pulsar (of a
''canonical'' surface magnetic field $B_{\rm i}\simeq 10^{12}$ G,
and a spin period of a fraction of a second) and a low-mass MS
companion. In a few million years, it spins down by means of magnetic dipole
braking down to the period of a few seconds, without suffering any
significant magnetic field decay. Consequently, it crosses the
radio pulsar death line and disappears as a pulsar. On a much
longer timescale of  a billion years, the low-mass companion enters
the red giant phase and fills its Roche lobe. This entails the mass
transfer onto the NS via an accretion disk. The binary system
becomes a LMXB, remaining in this stage for $10^8-10^9$~yrs.
Accretion of the plasma onto a NS increases its mass, while
accelerating its rotation, as well as inducing the decay
(dissipation) of its surface magnetic field.

The {\it accretion rate} is expressed in terms of the baryon mass
$M_{\rm b}$, since it is the parameter that can be uniquely determined
for the binary system; indeed $M_{\rm b}$ is related to the star's total
baryon number $N_{\rm b}$ by $M_{\rm b}=N_{\rm b}m_0$, where $m_0$
is the mass per nucleon of the $^{56}{\rm Fe}$ atom. The increase in 
$M_{\rm b}$ is straightforward and proportional to the accretion
rate - at time $t$ (measured by a distant observer) it is 
denoted as $\dot{M}_{\rm b}(t)$. Assuming that accretion
starts at $t_{\rm i}$, the integrated (total) baryon
mass increase is
\begin{equation}
\Delta M_{\rm b}(t)=
\int_{t_{\rm i}}^t \dot{M}_{\rm b}(t^\prime)
{\rm d}t^\prime.
\label{eq:DeltaMb.t}
\end{equation}
Since the detailed history of accretion is unknown, we treat
$\dot{M}_{\rm b}$ as a constant, such that
$\Delta M_{\rm b}(t)\approx \dot{M}_{\rm b}(t-t_{\rm i})$.
While $M_{\rm b}$ is a very important global stellar parameter,
the quantity that is actually measured is the {\it gravitational mass}
$M$. We define its increase as
\begin{equation}
\Delta M(t)=
\int_{t_{\rm i}}^t \dot{M}(t^\prime)
{\rm d}t^\prime.
\label{eq:DeltaM.t}
\end{equation}
It is found that $\Delta M$ depends on $\Delta M_{\rm b}$ according to the
relation between ${\rm d}M$, ${\rm d}M_{\rm b}$, and the total angular 
momentum $J$ (see e.g., \citealt{FIS})
\begin{equation}
 {\rm d}M=\Omega {\rm d}J + u^t{\rm d}\mb,
\label{smallinc}
\end{equation}
where $u^t$, the time component of fluid 4-velocity as seen by a distant
observer, has the meaning of chemical potential (divided by $m_0 c^2$).
For $\Delta M(t)$, one then obtains
\begin{equation}
\Delta M(t)=\int_{t_{\rm i}}^t
{\dot{M}_{\rm b}}(u^t + \ltot\Omega){\rm d}t^\prime,
\label{eq:Delta_M.Delta_M_b}
\end{equation}
using the total angular momentum $J$ evolution equation
\begin{equation}
\label{eq:evol0}
 \frac{{\rm d}J}{{\rm d}M_{\rm b}}=\ltot (\mb,f,B),
\end{equation}
where $\ltot$ includes the angular momentum transferred to the star by the
infalling matter and the influence of the magnetic torque (for details, see Sect.
\ref{sect:acc-B}). The values of the gravitational mass $M$ and spin frequency
$f=1/P$ for a given $M_{\rm b}$ and $J$ are calculated for stationary rigidly
rotating two-dimensional NS models \citep {BGSM}, with the {\tt rotstar} code implementation from the numerical relativity library 
LORENE\footnote{\tt http://www.lorene.obspm.fr}.
\section{Decay of neutron-star magnetic field in LMXBs}
\label{sect:decayB}

Millisecond pulsars have typically a low surface polar magnetic 
field of strength $B_{\rm p}\sim 10^8-10^9$~G (see e.g., \citealt
{Lorimer2008}), but observations do not provide any evidence of the 
$B_{\rm p}$ decay during the radio-pulsar phase. However, a 
substantial $B_{\rm p}$ decay (of some four orders of magnitude) is 
thought to occur during the accretion ''recycling'' in a LMXB, 
leading to the formation of a millisecond pulsar (\citealt
{Taam1986}, for a review see \citealt {Colpi2001}). We relate 
$B_{\rm p}$ to measured values of $P$ and $\dot{P}$ by a standard 
formula, $B_{\rm p} = 3.2\times 10^{19}\left(P\dot{P}/{\rm s}\right)^{1/2}$ G
\citep{ManchesterT1977}. This assumes a stellar moment of inertia 
$I=10^{45}~{\rm g\cdot cm^2}$ and radius $R=10$~km, and enables one 
to replace the measured value of $\dot{P}$ by $B_{\rm p}$.

\subsection{Decay of  magnetic field - theory}
\label{sect:decayB-theory}
Theoretical modeling of accretion-induced decay of $B_{\rm p}$ turns out to be a
challenging task. The original idea that $B_{\rm p}$ decays in close binaries because
it is `buried' (`screened') by the accreted matter was proposed by 
\citet{BisnovatyiKomberg1974}.
\citet{Romani1990} suggested that a combination of crustal heating due to accretion,
accelerating Ohmic dissipation, and advection of magnetic field lines from the
poles towards the equator could explain the $B_{\rm p}$ decay. Some other authors
focused on the realistic modeling of the acceleration of the Ohmic dissipation of the
crustal $B$ caused by the heating induced by accretion
(\citealt{GeppertUrpin1994,UrpinGeppert1995}). The decay of $B_{\rm p}$ due to the
diamagnetic screening by the accreted layer was studied in \cite{Zhang1994} and \citet{Zhang1998}. Dependence of the screening of crustal magnetic
field on the accretion rate was studied in more detail later by
\citet{Cumming2001}. These authors considered in detail the interplay between the
advection of magnetic field and its Ohmic diffusion, and used realistic microscopic
models of the outer `ocean' (molten crust) and the crust. Magneto-hydrodynamical simulation of accretion-burial of $B_{\rm p}$ have also been performed
by several authors (e.g. \citealt{PayneMelatos2004,PayneMelatos2007}, see also the study of \citealt{WetteVigelius2010} and references therein). This very brief
and incomplete (especially as far as the references are concerned)  review
illustrates the theoretical effort to explain the decay of $B_{\rm p}$ during the
accretion-driven recycling of the millisecond pulsars. Unfortunately, no reliable
and robust scenario based on realistic microphysics combined with complete
magneto-hydrodynamical treatment, and consistent with astronomical observations, is
available today. Therefore, in what follows we limit our description of the
$B_{\rm p}$ decay to simple phenomenological models, based to some extent on
observations of three populations: old radio pulsars, accreting binary neutron
stars, and millisecond pulsars. Our basic assumption (which is widely accepted) is  that these populations are linked by the recycling process, which in turn 
is driven by accretion onto a magnetized neutron star in a close binary.
\subsection{Decay of magnetic field - phenomenology (with some observational
basis)} \label{sect:decayB-phenom}
\citet{Taam1986} analyzed a set of LMXBs of different ages and 
therefore different amounts of accreted mass. They suggested that 
there is a possible inverse correlation between $B_{\rm p}$ and the 
(estimated) total amount of accreted material. Their conclusion was 
confirmed in a later study by \citet{vdHeuvel1995}. \citet{Shibazaki1989} 
presented more detailed arguments based on a subset 
of LMXBs that enabled this inverse correlation to be quantified as
\begin{equation}
B_{\rm p}=B_{\rm p}(\Delta M_{\rm b})=
B_{\rm i}/(1 +\Delta M_{\rm b}/ m_B),
\label{eq:B.DeltaM}
\end{equation}
where $B_{\rm i}$ is the initial (pre-accretion) magnetic field, $B_{\rm 
p}(\Delta M_{\rm b})$ is the magnetic field after $\Delta M_{\rm b}=\dot{M_{\rm b}}t$ is accreted by the neutron star, and  $m_{\rm B}$ is 
a constant setting the scale of dissipation of $B_{\rm p}$  with 
increasing $\Delta M_{\rm b}$. We note that \citet 
{Shibazaki1989} do not advocate any physical model leading to 
Eq.~(\ref{eq:B.DeltaM}). They report astrophysically interesting bounds on 
$m_{\rm B}$, resulting from the application of Eq.\;(\ref{eq:B.DeltaM}) 
to some LMXBs, X-ray pulsars, and millisecond pulsars. In particular, for 
$m_{\rm B}\ga 10^{-3}~\msun$, the evolution at an X-ray pulsar stage 
proceeds along the equilibrium spin-up line, with accretion 
(advection) torque balanced by the magnetic torque (\citealt
{GhoshLamb1979}, Fig. \;1 of \citealt{Shibazaki1989}) - rapid 
millisecond rotation cannot be obtained in this way. If, however, 
$m_{\rm B}\la 10^{-4}~\msun$, then the evolutionary tracks in the ${\rm 
log}B_{\rm p}-{\rm log}P$ plane are consistent with measured pairs 
of $B_{\rm p}$ and $P$ for binary and isolated millisecond radio pulsars. 
Finally, taking $m_{\rm B}\sim 10^{-4}~\msun$ allows a reasonable 
representation of the inverse correlation between $B_{\rm p}$ and 
$\Delta M_{\rm b}$ noted by \citet{Taam1986}.

One has to be aware that Eq.\;(\ref{eq:B.DeltaM}) is based on a
limited and uncertain set of data referring to LMXBs. It is clear
that Eq.\ (\ref{eq:B.DeltaM}) is too simplistic to describe
magnetic field decay for {\it all kinds} of accreting neutron stars.
It is  natural to expect that the decay of $B_{\rm p}$ depends not
only on $\Delta M_{\rm b}$, but also on $\dot{M}$, which is decisive in 
the heating of the neutron-star interior \citep
{Wijers1997,UrpinGeppert1996,UrpinGeppertKonenkov1998, Cumming2001}.
Specifically, \citet{Wijers1997} demonstrated that the decay law $B_{\rm
p}\propto 1/\Delta M_{\rm b}$ is inconsistent with a broader set of
available data for accreting neutron stars in both X-ray binaries and
recycled millisecond pulsars.

In spite of the limitations and uncertainties discussed above, 
we adopt Eq.\;(\ref{eq:B.DeltaM}) as our
baseline description of the $B_{\rm p}$ decay in LMXBs and compare
it with other proposed phenomenological formulae for the
accretion-induced $B_{\rm p}$ decay to assess
the model-independent features of our basic results.

For example, \citet{Wijers1997} discussed the square dependence on
$\Delta M_{\rm b}/ m_{\rm B}$ in the denominator of Eq.\;(\ref{eq:B.DeltaM})
as well:
\begin{equation}
B_{\rm p}=B_{\rm p}(\Delta M_{\rm b})=
B_{\rm i}/(1 +\Delta M_{\rm b}/ m_{\rm B})^2~.
\label{eq:B.DeltaM2}
\end{equation}
In this case, one has to use a lower value of $m_{\rm B} \simeq
10^{-3}\div 10^{-2}~\msun$, which allows to form a millisecond
pulsar before the field decays to very low values.

An exponential decay of $B_{\rm p}$ was employed by \citet{Kiel2008} 
and \citet{Oslowski2011} to be
\begin{equation}
B_{\rm p}=(B_{\rm i}-B_{\rm min})\exp(-\Delta M_{\rm b}/ 
m_{\rm B})+B_{\rm min},
\label{eq:Oslowski}
\end{equation}
where $B_{\rm min}=10^8$~G is the {\it assumed} minimal residual 
magnetic field, to reproduce the observed $P-\dot P$ distribution 
by means of the population synthesis studies \citep{Oslowski2011}. 
We discuss briefly how the final results depend on the assumed 
magnetic field decay model in Sect.~\ref{sect:results_fbm}.

\section{Spin-up by disk accretion onto a magnetized neutron star}
\label{sect:acc-B}
We assume that the evolution of an accreting NS can be represented
as a sequence of stationary rotating configurations of increasing
baryon mass. We use the KR formalism to determine the circular
orbit $r_0$, from where the accretion {\it effectively} takes place
- the radial inner boundary of the Keplerian accretion disk in
which the viscous torque is non-vanishing. Following KR, we define
the corotation radius $r_c \equiv \left(GM/\omega_s^2\right)^{1/3}$
(i.e., the radial distance at which the Keplerian orbital angular
frequency $\Omega_K$ is equal to the rotation frequency of a
central star, $\omega_s=2\pi f$), the magnetospheric radius
$r_m\equiv \left(GM\right)^{-1/7}\dot M^{-2/7} \mu^{4/7}$ (where
$\mu=B_{\rm p}R^3$ approximates the stellar magnetic dipole moment), the
magnetic-to-corotating radius ratio $\xi\equiv {r_m/r_c}$, and the
so-called ''fastness'' parameter $\omega\equiv
{\omega_s/\Omega_K(r_0)} = \left({r_0/r_c}\right)^{3/2}$. In
addition, we include relativistic effects - inevitable in the
space-time generated by a rotating NS - especially the existence of
the marginally stable orbit $\rms$ and denote the Schwarzschild
gravitational radius as $r_s \equiv {2GM/c^2}$, and the corotation
radius in $r_s$ units, $\beta\equiv r_c/r_s$.

We then solve the equation that corresponds to the vanishing 
of the viscous torque to determine the disk inner boundary, namely 
\begin{equation}
 \label{innerb}
\frac{{\rm d}l}{{\rm d}r}=-\frac{\mu^2}{\dot M r^4} (1-\omega^{-1}),
\end{equation}
where $l$ is the specific angular momentum of a particle calculated
in a relativistic way. 
To avoid the arduous calculations of the NS space-time that would be 
needed to obtain the specific angular momentum $l$ at a given
disk radius $r$ in every time-step of the evolution, we
exploit the result of \citet{BejgerZH2010}, who provide an
approximation of the value of $l(r)$, defined as
\begin{equation}
 l(r)=r\frac{v}{\sqrt{1-v^2/c^2}},
\label{ldef}
\end{equation}
with an approximate value of the particle orbital velocity $v$
\begin{equation}
 v={\frac{r}{\sqrt{1-2{GM/(rc^2)}}}}\left(\sqrt{\frac{GM}{r^3}}
-\frac{2GJ}{r^3c^2}\right).
\label{vdef}
\end{equation}
This approximate value of $l(r)$ deviates by less than one per
cent from the true value for spin frequencies and masses similar to
PSR J1903+0327 measurements (for details see \citealt{BejgerZH2010}).
Eqs.~(\ref{innerb}-\ref{vdef}) yield an algebraic equation for $r_0$ of 
\begin{equation}
\label{bc}
\frac{1}{2} f_{\rm ms}(r_0) = \left(\frac{r_m}{r_0}\right)^{7/2} \!\! \left(\sqrt{\frac{r_c^3}{r_0^3}}-1\right) =
\frac{\xi^{7/2}}{\omega^{10/3}}\left(1-\omega\right), 
\end{equation}
with a dimensionless function $f_{\rm ms}$ defined as 
\begin{equation}
f_{\rm ms}(r)=\frac{2}{\Omega r} \frac{{\rm d}l}{{\rm d}r}, 
\label{fmsdef}
\end{equation}
which for $l(r)$ given by Eqs.~(\ref{ldef}) and (\ref{vdef}) equals
\begin{eqnarray}
 \label{frel}
f_{\rm ms}(r_0)&=&\frac{1-\alpha/\rb^{3/2}}{(1-v^2/c^2)^{3/2}\sqrt{1-1/\rb}}\times \nonumber \\ 
&\times&\left(\frac{\rb-2}{\rb-1}-2\frac{v^2}{c^2}+\frac{3\alpha}{\rb^{3/2}-\alpha}\right),
\end{eqnarray}
where $\alpha=Jc/(\sqrt{2}GM^2)$ and
$\rb=r_0/r_{s}=\beta\omega^{2/3}$.

In general, Eqs.~(\ref{bc}) and (\ref{frel}) allow us to calculate $r_0$ in   
the ''recycling'' process of even very rapidly rotating and massive 
millisecond pulsars, when the existence of a relativistic 
marginally-stable orbit cannot be neglected i.e., when the KR condition of 
$r_0 \ll r_{\rm ms}$ is no longer valid. The value of $r_{\rm ms}$ corresponds 
to the solution of
\begin{equation}
 \label{msorb}
\frac{{\rm d}l}{{\rm d}r}=0, 
\end{equation} 
which is determined by assuming that $f_{\rm ms}(r_{\rm ms})=0$. The validity of employing 
this general, refined approach in the case of PSR J1903+0327 is studied in the
Appendix.

To calculate the increase in the total stellar angular momentum
$J$, we take into account the transfer of specific angular momentum
$l_0\equiv l(r_0)$ and the KR prescription for the magnetic torque 
\begin{equation}
\label{eq:evol}
 \frac{{\rm d}J}{{\rm d}\mb}=\ltot=l(r_0) -\frac{\mu^2}{9r_0^3 {\dot\mb}}
 \left(3-2\sqrt{\frac{r_c^3}{r_0^3}}\right).
\end{equation}
\section{Equations of state}
\label{sect:EOS}
The EOS of dense cores of NSs remains poorly constrained. This 
is due to, on the one hand, a lack of knowledge of strong interactions in dense matter,
and on the other hand, deficiencies in the available many-body
theories of dense matter. This uncertainty has been reflected
as a rather broad scatter in the theoretically derived and EOS dependent
maximum allowable masses for NSs, $M_{\rm max}({\rm EOS})$ (see, e.g.
\citealt{NSbook2007}). Fortunately, the measurement of the mass of
PSR J1614-2230, of $1.97\pm 0.04~{\rm M}_\odot$ \citep{Demorest2010},
introduces a rather strong constraint of $M_{\rm max}\ge2.0~\msol$. This
means that the true EOS is rather stiff. To illustrate a remaining 
uncertainty in the stiffness, we considered three different
models of the EOS. In all cases, the simplest composition of matter was 
assumed - neutrons, protons, electrons, and muons in $\beta$-equilibrium
($npe\mu)$:
\begin{list}{\labelitemi}{\leftmargin=1.5em}
\item[{\bf DH}] This model of \citet{DouchinH2001} is non-relativistic and
its energy density functional is based on the SLy4 effective nuclear
interaction. The model describes in a unified way both the dense
liquid core of NS and its crust, yielding $M_{\rm max}=2.05~\msol$ and 
 a circumferential radius at the maximum mass of $R_{M_{\rm max}}=10.0~$km.
\item[{\bf APR}]$={A}18+\delta v +{\rm UIX}^*$ from \citet{AkmalPR1998} is a 
non-relativistic model with some relativistic corrections.
It consists of a two-nucleon Argonne potential A18 with relativistic boost
corrections $\delta v$  and an adjusted three-nucleon Urbana UIX*
potential. A variational solution of the many-body problem 
yields $M_{\rm max}=2.21~\msol$ and a circumferential radius at the 
maximum mass of $R_{M_{\rm max}}=10.0~$km.
\item[{\bf BM}]$=TM16S0$ (with some minor changes) was drawn from a set of 
relativistic models of \citet{BednarekM2009}. It consists of a Lorentz-covariant 
effective nonlinear Lagrangian including up to quartic terms in meson fields,
based on chiral-symmetry breaking expansions. The EOS was calculated in the
mean field approximation, yielding $M_{\rm max}=2.11~\msol$ and a 
circumferential radius at the maximum mass of $R_{M_{\rm max}}=11.95~$km.
\end{list}

\section{Results}
\label{sect:results}
To illustrate the laws and relations governing the
,,recycling'' process, we begin by constraining the final
spin frequency, $f=465$ Hz. We then calculate sets of the
evolutionary tracks labeled with the NS initial parameters - $M_{\rm i}$,
$P_{\rm i}$, and $B_{\rm i}$ - covering a broad range of possible
values. If not stated otherwise, the figures relate to results obtained
using the DH EOS. In what follows, we denote $B_{\rm p}$ by $B$ 
and $\dot{M}_{\rm b}$ by $\dot{M}$.

\subsection{Constraining the parameter space: final frequency $f=465$ Hz}
\label{sect:finfreq}
\begin{figure}
\resizebox{\hsize}{!}{\includegraphics[]{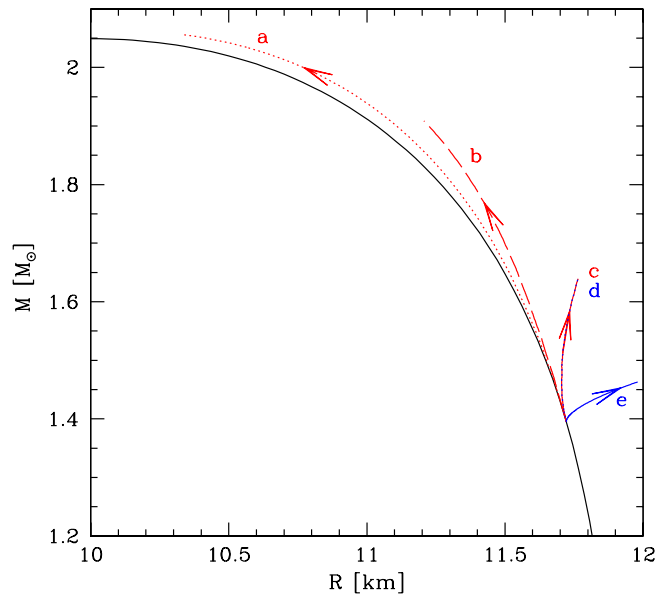}}

\caption{Mass-radius relation for accreting stars with different
initial magnetic field $B_i$ and accretion rate $\dot M$. Solid black 
curve denotes static configurations. Evolutionary tracks (arrows mark 
the direction of evolution) correspond to 
the following initial parameters ($B_i$~[G], $\dot M~[\msolyr]$) -
{\bf a}: ($10^{12}$, $3\times 10^{-11}$),
{\bf b}: ($10^{12}$, $10^{-10}$),
{\bf c}: ($10^{12}$, $10^{-9}$),
{\bf d}: ($10^{11}$, $10^{-11}$),
 and {\bf e}: ($10^{11}$, $10^{-9}$).
Tracks {\bf c} and {\bf d} coincide, as explained in
Sect. \ref{sect:finfreq} (color online).}
\label{fig:mr24}
\end{figure}
\begin{figure}
\resizebox{\hsize}{!}{\includegraphics[]{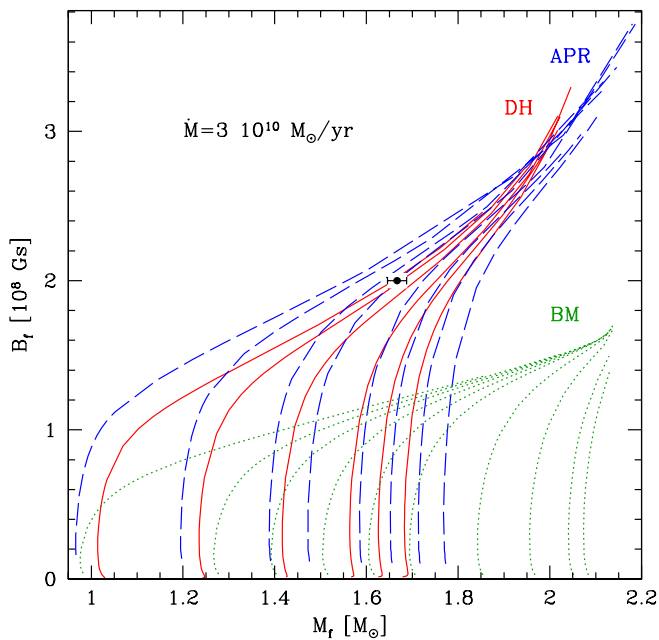}}
\caption{Final magnetic field $B_f$ vs the gravitational mass
for a star rotating at $f=465$ Hz. Different colors correspond
to different EOSs (solid red - DH, dashed blue - APR, dotted green - BM).
The accretion rate is $\dot M=3\times 10^{-10}\ \msolyr$.
The error bar reflects the uncertainty ($3\sigma$) in
the mass measurement (color online).}
\label{fig:bfmfeos}
\end{figure}

Fig. \ref{fig:mr24} presents the mass-radius relation
for accreting NSs for different values of accretion rate $\dot M$
and initial magnetic field $B_i$. The initial mass is fixed at
$1.4\ \msol$. The upper ends of each curve correspond to the final
frequency, $465$~Hz. For $B_i=10^{12}$ G, the lowest accretion
rate considered is $3\times  10^{-11}~\msolyr$ because for lower
rates the configurations enter the axisymmetric-perturbation
instability region already for frequencies lower than $465$ Hz.
\begin{figure}
\resizebox{\hsize}{!}{\includegraphics[]{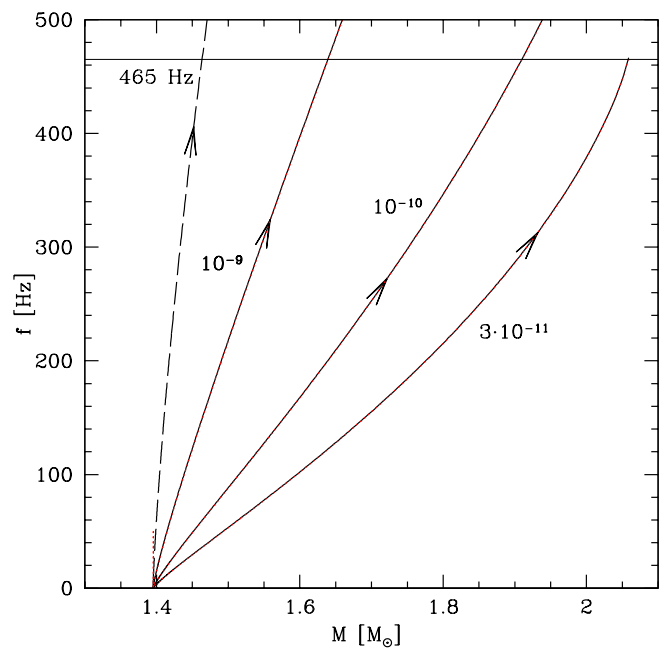}}

\caption{Spin frequency evolution during accretion, for initial
frequencies zero and 50 Hz, as a function of stellar mass. Three
different cases of accretion rates (in $\msol/{\rm yr}$)
 for initial magnetic field
$B_i=10^{12}$ G (solid line), $10^{11}$ G (dashed line) are shown
(color online).}
\label{fig:fmsl14}
\end{figure}
\begin{figure}
\resizebox{\hsize}{!}{\includegraphics[]{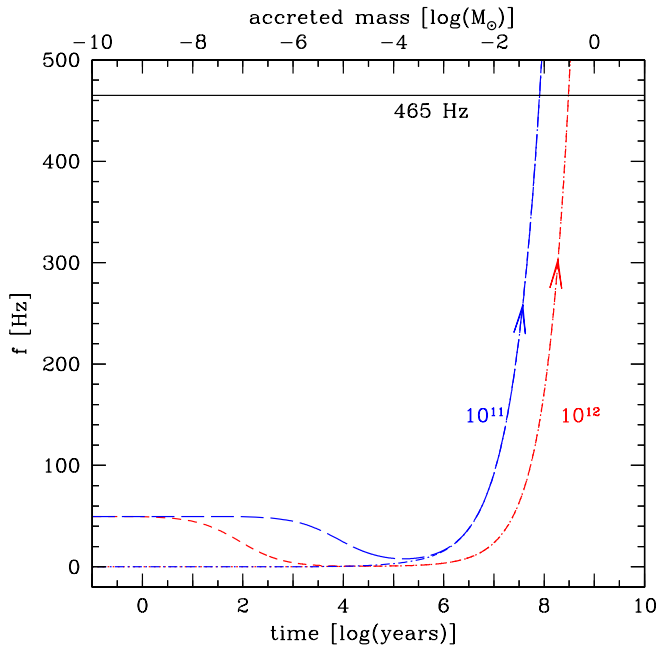}}
\caption{Spin frequency evolution for initial spin
frequencies of zero and 50 Hz and for an initial magnetic field of 
$B_i=10^{12}$ G and $10^{11}$ G as a function of the accreted mass
(upper axis) or time (lower axis) calculated for a constant accretion
rate $\dot M=10^{-9}~\msolyr$, on a logarithmic scale (color online).}
\label{fig:ftslg}
\end{figure}
The same configurations are shown in Fig. \ref{fig:fmsl14}, where the
rotational frequency as a function of mass is
presented for different accretion rates. For some evolutionary
tracks in Fig. \ref{fig:fmsl14}, the initial frequency is set to
$f_i=50$~Hz, the value corresponding to a typical frequency
expected for newly-born pulsars (Table 7 of \citealt{faucher}).
These curves are indistinguishable from those corresponding to
initially non-rotating configurations, as the braking timescale
is by many orders of magnitude shorter than the spin-up time.
The difference is visible on a logarithmic scale in Fig. \ref{fig:ftslg},
where we present the frequency evolution of an accreting star
for two initial configurations: non-rotating NS and
a configuration rotating initially with frequency 50 Hz.
The accretion rate here is fixed to $\dot{M} = 10^{-9}~\msolyr$.
For $B_{\rm i}\sim 10^{12}$~G, the {\it spin-down} predominates for
$\sim 10^3$~yrs, and afterwards the accretion {\it spins-up} the star.
The amount of accreted material depends very sensitively on the strength
of the initial magnetic field; correspondingly, for $B_i=10^{11}$~G the spin-down
timescale is two orders of magnitude longer ($10^5$~yrs) than for
$B_i=10^{12}$~G. This is a direct consequence of the quadratic
dependence of the magnetic torque on the magnetic field $B$,
Eq. (\ref{eq:evol}). The minimum value of rotation frequency
corresponds to the exact balancing of angular momentum $l_0(r_0)$
at the accretion disk edge, $r_0$, by the magnetic torque.
\begin{figure}
\resizebox{\hsize}{!}{\includegraphics[]{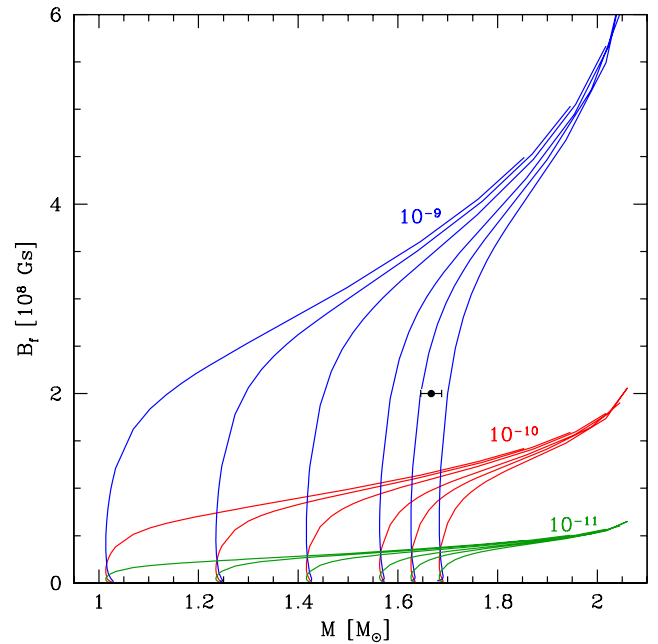}}
\caption{Final magnetic field $B_f$ versus gravitational mass $M$
for configurations rotating at $f=465$ Hz. Colors correspond
to different accretion rates. For a given accretion rate
(in $\msol/{\rm yr}$), 
different curves are defined by a given initial configuration
(i.e., central density or initial mass). Along each curve, the 
initial magnetic field increases upwards. The error bar corresponds
to the uncertainty ($3\sigma$) in the mass measurement 
(color online).
}
 \label{fig:bfmf}
\end{figure}
In Figs. \ref{fig:bfmf} and \ref{fig:bfmfeos}, we show
the magnetic field vs. mass dependence in such a case (the 
error bar corresponds to the $3-\sigma$ measurement of the mass,
$M=1.667\pm 0.021~\msol$).

We note that the problem under consideration is degenerate with 
respect to the constant
\begin{equation}
q\equiv B^2/\dot M ,
\label{eq:q}
\end{equation}
(see Fig. \ref{fig:bfmf}), as a consequence of
Eqs. (\ref{bc}) and (\ref{eq:evol}) depending on the
quantity $q$, and not $B$ and $\dot M$ separately.

To obtain the results for some $\dot M_2$, the magnetic
field $B$ corresponding to $\dot M_1$ should be multiplied by the
factor $\sqrt{{\dot M_2} /{\dot M_1}}$.  This relation allows us to
determine the lower bound to the accretion rate for given observed
values of $B$ and $M$ or, assuming  an accretion rate, to determine the
maximum value of the final magnetic field e.g., if 
$\dot{M}=\dot{M}_{10}=10^{-10}~\msolyr$ 
(red curve on Fig. \ref{fig:bfmf}), $B_{\rm max} =
1.16\times 10^8$ G for $M=1.67~\msol$. From the observations, we infer that 
$B_f\simeq 2\times 10^8$ G; since the condition $B_{\rm max}>B_f$ should
be fulfilled, we use the scaling law 
\begin{equation}
B_{max} (\dot M)=B_{max} (\dot{M}_{10})
\left(\frac{\dot{M}}{\dot{M}_{10}}\right)^{1/2},
\label{eq:bmdot}
\end{equation}
to obtain a lower limit to the accretion rate
\begin{equation}
\dot M > \left(\frac{B_f}{B_{max} (\dot{M}_{10})}\right)^2 \dot{M}_{10} 
= 3\times 10^{-10}~\msolyr.
\label{eq:mdotmin}
\end{equation}
In Fig. \ref{fig:bfmfeos}, we present the relation between the
magnetic field and the gravitational stellar mass for the
configuration spun up to $465$~Hz, for three different EOSs
described in Sect. \ref{sect:EOS}. The lower limit to $\dot{M}$
assessed above is marginally consistent with the DH EOS for the
measured values of $M=1.67~\msol$ and $B=2\times 10^8$~G.

\subsection{Constraining the parameter space even further:
$f=465$ Hz and $M=1.67~\msol$}
\begin{figure}
\resizebox{\hsize}{!}{\includegraphics[]{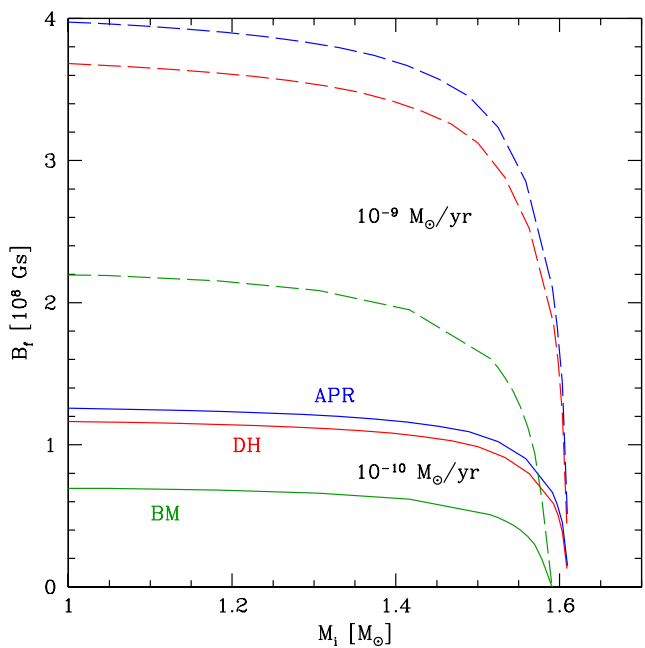}}
\caption{Final magnetic field, $B_f$, versus initial gravitational mass,
for a star with final parameters $M=1.67~\msol$ and the frequency $f=465$ Hz,
for APR, DH, and BM EOSs. The accretion rate is
$\dot M=10^{-10}\ \msolyr$ (solid curves) and
$\dot M=10^{-9}\ \msolyr$ (dashed curves). Scaling relation
$B_f (\dot M_1) = B_f (\dot M_2) \cdot \sqrt{\dot M_1/ \dot M_2}$
is fulfilled, i.e., dashed curves match the solid ones
multiplied by $\sqrt{10}$ (color online).}
\label{fig:mibf}
\end{figure}

Lines presented in Fig. \ref{fig:mibf} correspond to different
initial parameters (magnetic field $B_i$ and mass $M_i$) that lead
to configurations of $M=1.67\ \msol$ at the spin frequency $f=465$ Hz.
We employ two fixed accretion rates of $\dot M=10^{-10}\ \msolyr$ and
$\dot M=10^{-9}\ \msolyr$. The scaling relation $B_f (\dot M_1) = B_f
(\dot M_2) \cdot \sqrt{\dot M_1/ \dot M_2}$ holds and allows us to
set limits on $\dot M$ for a given value of the magnetic field.

The differences between the DH and APR EOSs are small in comparison to
the difference of them both with a much stiffer BM EOS. The value of the 
stellar moment of inertia is responsible for such a discrepancy: $I(M)$ for the BM
EOS is about 25\% higher than the corresponding values for the DH
and APR EOS (for the same $M$). Consequently, to obtain
the same final frequency one needs a larger $J$ and a larger amount
of accreted mass, which leads to the lower value of $B_f$.

\subsection{From a progenitor NS to PSR J1903+0327:
$M=1.67~\msol$, $f=465$ Hz, and $B=2\times 10^8$~G}
\label{sect:results_fbm}
One can obtain a stringent limit on the parameters of the
progenitor NS by fixing the  final value of the
magnetic field.
\begin{figure}
\resizebox{\hsize}{!}{\includegraphics[]{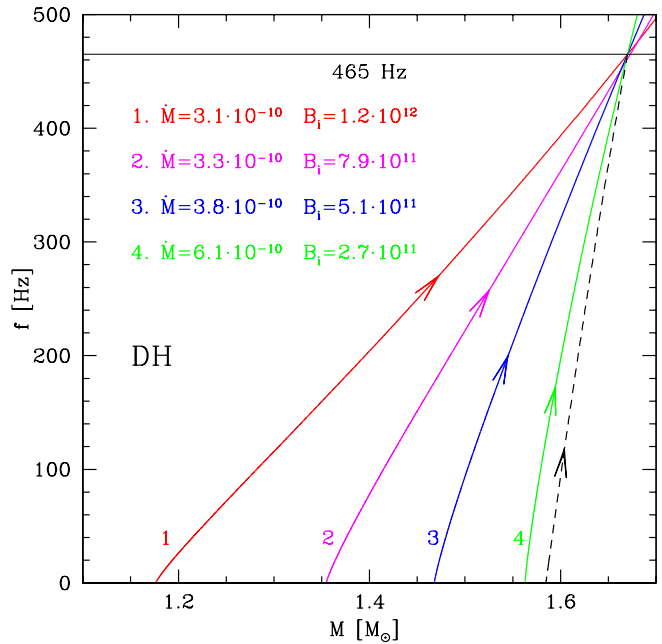}}
\caption{Spin-up tracks of the accreting NSs leading to
the final configuration rotating at $f=465$ Hz and with gravitational
mass $M=1.67~\msol$ and magnetic field $B=2\times10^8$~G. Curves are
labeled by the
average accretion rate (in $\msol/{\rm yr}$ and initial value of the
magnetic field (in G).
For comparison the spin-up for $B=0$ (dashed line), via accretion
from marginally stable orbit, is shown (color online).}
 \label{fig:fmsl}
\end{figure}
Some tracks considered in this subsection are shown
in Fig. \ref{fig:fmsl}, labeled by the initial value of
the magnetic field $B_i$ and the accretion rate $\dot M$.
The final magnetic field is fixed to $B=2\times10^8$~G.
For comparison, we present the case in which we neglect the effect
of the magnetic field entirely: the accretion takes place from the
marginally-stable orbit calculated exactly in accordance with 
General Relativity (see, e.g., \citealt{Zdunik2002}).

The spin-up in the $B=0$ case is very efficient, one needs $\simeq 
0.1~\msun$ only to reach the observed frequency of $465$ Hz. 
Neglecting a magnetic field therefore leads to a rather high 
progenitor-NS mass, $1.58\ M_\odot$, whereas reasonable values of 
the progenitor $B$ and $\dot{M}$ enable $M_f$ to be reached starting 
from a moderate value of $M_i \sim 1.3\ M_\odot$. The amount of 
accreted matter in the $B=0$ case can be larger if we allow for the 
reduction in the angular momentum transport efficiency (parameter 
$x_l\ll 1$ in \citealt{Zdunik2002}), here however we assume a 100\% 
efficiency ($x_l\equiv 1$), in accordance with recent numerical 
calculations \citep{BeckwithHK2008,ShafeeMNTGM2008}.

\begin{figure}
\resizebox{\hsize}{!}{\includegraphics[]{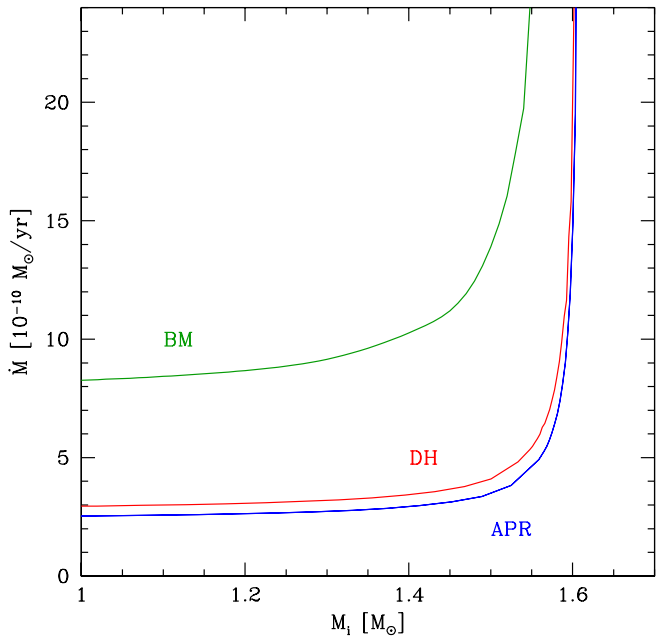}}
\caption{Average accretion rate vs initial mass  needed to evolve
a  NS by disk accretion to the frequency $f=465$ Hz, mass
$M=1.67~\msol$, and magnetic field $B=2\times 10^8$~G for the 
three EOSs under consideration (color online).}
 \label{fig:midma}
\end{figure}

\begin{figure}
\resizebox{\hsize}{!}{\includegraphics[]{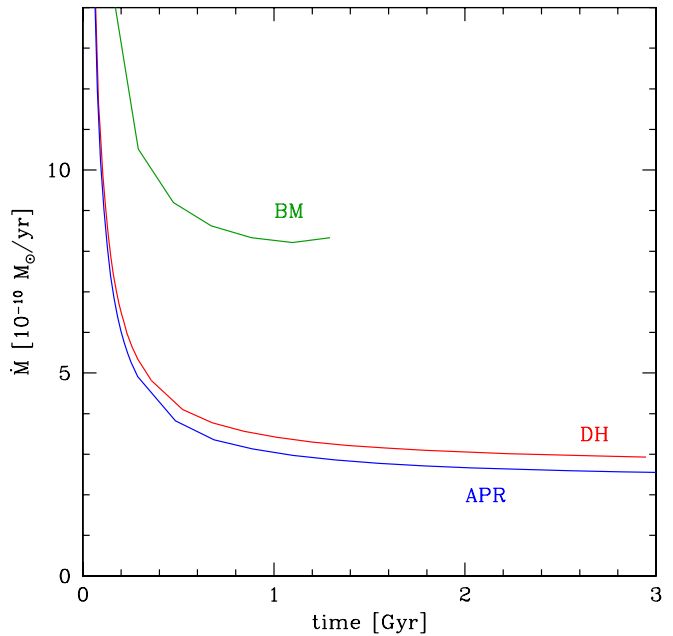}}
\caption{Accretion rate versus time of accretion needed to
reach observable configuration (color online).}
\label{fig:tdma}
\end{figure}

The accretion rate as a function of $M_i$ (assuming final $B_f=2\times10^8$~G)
is presented in Fig. \ref{fig:midma}.  The lower limit to the accretion rate
depends on the EOS and ranges from $2.5\times 10^{-10}\ \msolyr$ (DH, APR) to
$8.5~\times 10^{-10}\msolyr$ (BM). For a given EOS, the required $\dot M$
depends weakly on the accreted mass needed to spin up the star to 465 Hz,
provided that $\Delta M >0.2~\msol$.  For  $\Delta M < 0.1~\msol$, the required
accretion rate increases dramatically with decreasing $\Delta M$.  Using
$\Delta\mb=\dot\mb t$, we obtain the time needed to spin up the star to its
presently observed frequency. The result is presented in Fig. \ref{fig:tdma}.

We also performed calculations for other $B(\Delta M)$ dependences (Eqs.
\ref{eq:B.DeltaM2} and \ref {eq:Oslowski} in Sect.~\ref{sect:decayB}). The
results are qualitatively similar in the case of an inverse quadratic dependence
(Eq. \ref {eq:B.DeltaM2}); the value of $\dot M$ at the ''plateau'' region in
Figs.\ \ref{fig:midma} and \ref {fig:tdma} is slightly higher, $3.6\times
10^{-10}\ \msolyr$, than $3\times 10^{-10}\ \msolyr$, while the Eq.\
(\ref{eq:B.DeltaM}) is adopted. For the exponential field decay (Eq.\
\ref{eq:Oslowski}) however, the resulting $\dot M$ is $8\times 10^{-10}~\msolyr$
(for the mass decay scale $m_B = 0.05~\msun$, which according to
\citeauthor{Oslowski2011}, gives the closest agreement between their population
synthesis model predictions and the observed $P-\dot P$ data for Galactic pulsars). This
discrepancy is due to the different behavior of $B_{\rm p}(\Delta M)$ close to
the final $B_{\rm p}$ value, where Eq.~(\ref{eq:Oslowski}) gives significantly
larger values of $B_{\rm p}$ than Eq.~(\ref {eq:B.DeltaM}). For an efficient
spin-up, the ''magnetic term'' proportional to $B_{\rm p}^2/\dot M$ in $l_{\rm
tot}$ should be small (Eq. \ref{eq:evol}), hence the relatively large value of
$B_{\rm p}$ has to be compensated for by a higher value of $\dot M$ - for example,
assuming $B_{\rm i}=10^{12}$~G, Eq.~(\ref {eq:B.DeltaM}) yields $\Delta M =
0.5\ \msun$ in $1.7\times 10^9$ ~yrs, while the exponential decay of
\citet{Oslowski2011} gives $0.46\ \msun$ in $6\times 10^8$~yrs, that is, 
the accretion of almost the same amount of matter but somewhat more rapidly.
One should however treat this kind of empirical formulae with caution, since in the 
case of parameters obtained by population synthesis methods they may be uncontrollably 
affected by other assumptions.

\section{Conclusions}
\label{sect:conclusions}

Our simulations have allowed to estimate the intrinsic parameters of the progenitor NS
required to reach, in the process of recycling,  the measured  parameters of PSR
J1903+0327. To some extent, as we have shown, the  progenitor NS parameters depend
on the EOS of dense matter. This was studied using  three EOSs consistent with the 
measurement of $2\msol$ NS \citep{Demorest2010}.

We have found that the mean accretion rate $\dot M$ during recycling should be larger
than $(2.5 \div 8.5)\times 10^{-10}~\msolyr$, the highest lower bound being 
obtained for the stiffest EOS.

For each EOS, the required mean accretion rate is approximately constant for a broad
range of initial masses, $1~\msol\div 1.4~\msol$.  Therefore, depending on the
initial magnetic field $B_i$ we can reproduce parameters of PSR J1903+0327, or
more generally, observed millisecond pulsars, for a specific range of initial
masses. In other words, the present parameters of a recycled millisecond pulsar do
not allow us to determine its initial mass, in contrast to the case of recycling by
accretion with $B=0$. Simulations that neglect the magnetic field (following
the seminal paper of \citealt {CookST1994}) give a rather high value of the
lower bound to the progenitor NS mass of $M_i>1.55~\msol$ ($1.58~\msol$ for DH
and APR EOS, and $1.55~\msol$ for the BM EOS). This is a direct consequence of
the finding that the magnetic field significantly {\it decreases} the spin-up rate
(effectively {\it decreasing} the efficiency of angular momentum transfer onto
the star, as in to the $B=0$ {\it and}  $x_l\ll 1$ case). Accounting for
the magnetic field effect (magnetic torque) is therefore crucial for the
understanding of the observable pulsar population properties, especially in view of
the proposal that gravitational wave emission is a major dissipative agent that prevents
efficient spin-up \citep{ArrasFMSTW2003,WattsKBS2008,WattsK2009}. In other
words, for evolutionary reasons there may not be sufficient time {\it and/or} a
sufficient amount of matter to accrete, to form a
rapidly-spinning pulsar with a magnetic field sufficiently strong to produce a
detectable radio beam.

The framework presented here is suitable for testing the global properties of
the Galactic pulsar population, as well as for studying other millisecond
pulsars with precisely measured masses, e.g., massive millisecond pulsars
recycled in intermediate-mass X-ray binaries, where the recycling process is
much shorter than the one studied here \citep{Tauris2011}.  Our analysis can
also be easily extended to take into account time-varying accretion rates and
more detailed multi-parameter magnetic field decay prescriptions - this paper
is the first of a series devoted to these studies.

\begin{figure}
\resizebox{\hsize}{!}{\includegraphics[clip]{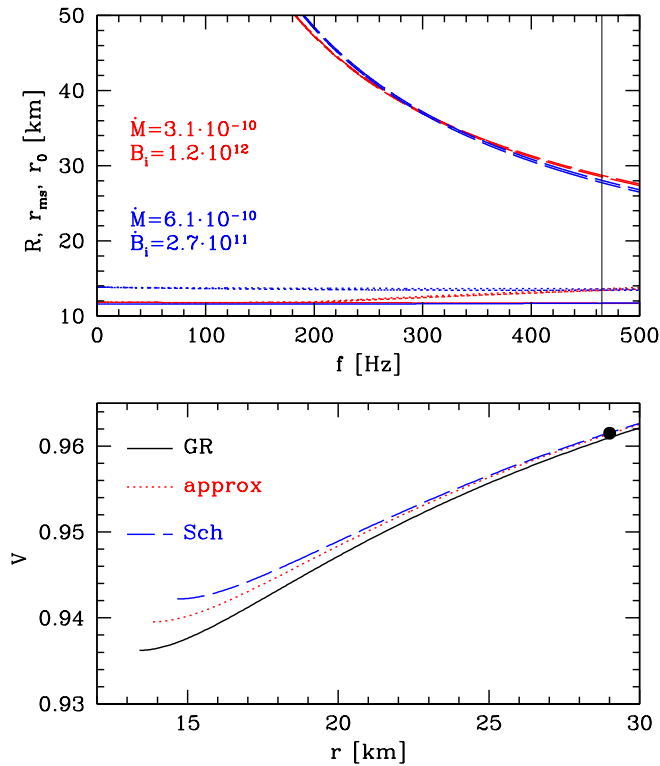}}
\caption{Upper panel: spin-up tracks of accreting NSs leading to presently
observed PSR J1903+0327 parameters on radius-spin frequency plane. 
The relevant characteristic radii are: stellar radius (solid curves), 
the radius of a marginally stable orbit (dotted curves), and the radius 
of the inner boundary of accretion disk (dashed curves). Colors correspond 
to the specific values of accretion rate and the initial value of the magnetic 
field. Lower panel: effective radial potential $V\left(r,l(r)\right)$ in 
the energy ($mc^2$) units for an exact, approximate, and Schwarzschild solution; 
the dot marks the final $r_0$. See the text for more details (color online).}
\label{fig:rfsl}
\end{figure}

\acknowledgements{ We are grateful to W. Klu{\'z}niak for his helpful comments
referring to the Klu{\'z}niak-Rappaport model of magnetically-torqued accretion
disks. We also acknowledge the helpful remarks of participants of the CompStar
2011 Workshop (Catania, Italy, 9-12 May, 2011), after the talk by one of the
authors (JLZ). This work was partially supported by the Polish MNiSW grant no. 
N N203 512 838, by the LEA Astrophysics
Poland-France (Astro-PF) program, and the ESF Research Networking Programme 
CompStar. MB acknowledges Marie Curie Fellowship within the 7th European 
Community Framework Programme (ERG-2007-224793).}

\section*{Appendix: The importance of relativistic effects in the
''recycling'' process of PSR J1903+0327}
\label{sect:appendix}

The original framework of \cite{klurap} is restricted to the non-relativistic
limit, i.e., it does not include the effects of the existence of a marginally
stable orbit of radius $\rms$.  As mentioned by these authors, their model
should be used only where $r_0 \gg \rms$. To establish a more general model
that can be applied to compact and massive configurations near the
mass-shedding limit, we have amended this shortcoming here 
by introducing Eq. (\ref{bc}). We demonstrate in this Appendix however, 
that in the case of moderately fast-spinning
PSR J1903+0327 this refined approach is not essential.

The upper panel of Fig. \ref{fig:rfsl} shows three characteristic 
radii of the problem, $r_0$, stellar radius $R$, and $\rms$, 
calculated along a spin-up evolutionary track. For the final 
configuration, $f=465$ Hz, $M=1.67~\msol$, and $B=2\times 10^8$~G, 
where $r_0$ reaches its minimum value, $r_0/\rms\simeq 2$. We 
compare two results, one where the effect of General Relativity is 
mitigated by Eq.\ (\ref{bc}), and the original model of KR, $f_{\rm 
ms}=1$ in Eq.\ (\ref{bc}). The difference in $r_0$ values is $\sim 
200$~m, i.e., $0.7\%$. Lower panel shows the effective radial 
potential $V\left(r,l(r)\right)$ ($l(r)$ is the particle specific 
angular momentum on the circular orbit of the radius $r$; see, e.g., 
Eq.~2 of \citealt{BejgerZH2010}); the final $r_0$ is sufficiently 
far from the $\rms$, hence we conclude that in the particular case 
of PSR J1903+0327 the relativistic correction necessary to account 
for $r_{\rm ms}$ can be omitted.

%
\end{document}